\documentstyle[preprint,tighten,eqsecnum,floats,epsfig,aps]{revtex}

\newcommand{\beq}{\begin{equation}}
\newcommand{\eeq}{\end{equation}}
\newcommand{\mc}{\multicolumn}
\newcommand{\lsim}{\mathrel{\mathop{\kern 0pt \rlap
  {\raise.2ex\hbox{$<$}}}
  \lower.9ex\hbox{\kern-.190em $\sim$}}}
\newcommand{\gsim}{\mathrel{\mathop{\kern 0pt \rlap
  {\raise.2ex\hbox{$>$}}}
  \lower.9ex\hbox{\kern-.190em $\sim$}}}

\begin{document}
\draft

\title{Local Realism for $K^0\bar{K}^0$ pairs}
\author{G. Garbarino}
\address{Grup de F\'{\i}sica Te\`{o}rica, 
Universitat Aut\`{o}noma de Barcelona, \\
E--08193 Bellaterra (Barcelona), Spain\\ and \\
Departament d'Estructura i Constituents de la Mat\`{e}ria, \\
Universitat de Barcelona, E--08028 Barcelona, Spain}
\date{\today}
\maketitle

\begin{abstract}
In this talk we discuss the predictions of local realistic theories for
the evolution of a $K^0\bar{K}^0$ quantum entangled
pair created in the decay of the $\phi$--meson. 
It is shown, in agreement with Bell's theorem,
that the most general local hidden--variable model fails in
reproducing the whole set of quantum--mechanical observables.
We achieve this conclusion by employing two different approaches.
In the first approach, the local realistic observables are deduced
from the most general premises concerning locality and realism,
and Bell--like inequalities are not employed.
The other approach makes use of Bell's inequalities.
Under particular conditions for the detection times,
within the first approach the discrepancy between
quantum mechanics and local realism for the asymmetry
parameter turns out to be not less than 20\%.
A similar incompatibility can be made evident by means of a Bell--type test,
by employing a Clauser--Horne--Shimony--Holt's inequality written
in terms of properly normalized observables.
Because of its relatively low experimental accuracy,
the data obtained by the CPLEAR collaboration
do not yet allow a decisive test
of local realism. Such a test,
both with and without the use of Bell's inequalities, should be 
feasible in the future at the Frascati $\Phi$--factory.

\end{abstract}

%\newpage
\pagestyle{plain}
\baselineskip 16pt
\vskip 48pt

%\newpage
%\doublespace
%*****************************************************************
% 1. Hystorical Introduction
%*****************************************************************
\section{Hystorical Introduction}
\label{intro}
In 1935 Einstein Podolsky and Rosen (EPR in the following) \cite{Ei35} advanced a 
strong criticism concerning the interpretation of quantum theory.
Starting from a few premises concerning {\it completeness}, {\em physicsl
reality} and {\it locality} and
considering the behaviour of a correlated and non--interacting
system composed by two separated entities, EPR arrived at the conclusion
that the description of reality given by Copenhagen's interpretation is
incomplete. 
At the very heart of their logical conclusion is the following
fact: their assumption, that a quantum system has real and well defined
properties also when does not interact with other systems 
(including a measuring apparatus), is contradicted by quantum mechanics. 

This was the point attacked by Bohr in his famous response \cite{Bo35} to EPR's 
paper. He noticed that EPR's criterion of
reality contained an ambiguity if applied to quantum phenomena.
Starting from the complementarity point of view, Bohr stressed again that {\it quantum
mechanics within its scope {\rm [namely, in its form restricted to human knowledge]}
would appear as a completely rational description of the physical phenomena}.

The first hypothesis for the solution of the paradoxical conclusion of EPR
was proposed by Furry \cite{Fu36} in 1935. He assumed that
in presence of EPR correlations between two quantum subsystems
which are very far away one from each other, the state
of the global system is no longer given by a superposition
of tensorial products of states but it is simply 
represented by a statistical mixture of products of states.
However, different experiments
excluded a possible separability of the
many--body wave function even in the case of space--like separated particles. 

EPR's paradox was interpreted as the need for the introduction,
in quantum theory, of additional variables, in order to restore
{\it completeness}, {\it relativistic causality} (namely {\it locality})
and {\it realism}.
In 1952 Bohm \cite{Bo52} suggested an interpretation of quantum theory
in terms of {\it hidden--variables}, in which the general mathematical
formulation and the empirical results of the theory remained unchanged. 
In Bohm's interpretation the paradoxical behaviour of correlated
and non--interacting systems revealed by EPR find an explanation. 
However, for such systems Bohm's theory exhibits a non--local character.

This result is consistent with what Bell obtained in 
1964 \cite{Be64}. He proved that any {\it deterministic local
hidden--variable theory} is incompatible with some statistical
prediction of quantum mechanics. This is the content of Bell's theorem
in its original form, which has been generalized in \cite{Be71} 
to include {\it non--deterministic theories}. From then, Bell and other 
authors \cite{CHSH69,Wi70,CH74,Cl78} derived different inequalities suitable for
testing what has been called {\it local realism}. 

Once established the particularity of Bell's local
realism, different experiments have been 
carried out to test these theories. The oldest ones \cite{Cl78,As82} measured the
linear polarization correlations of photon pairs created in radiative
atomic cascade reactions or in electron--positron annihilations, 
whereas, more recently, parametric
down--conversion photon sources have been employed \cite{Ze99,Ti98,We98}. 
Essentially all the experiments performed until now (in optics and atomic physics)
have proved that the class of theories governed by Bell's 
theorem are unphysical.
Actually, to be precise, because of apparata non--idealities
and other technical problems,
supplementary assumptions are needed in the interpretation of the experiments,
and, consequently, no test employed to refute local realism
has been completely loophole free \cite{Cl78,Ti98}.

It is then important to continue performing experiments on 
correlation properties of many particle systems, possibly in new
sectors, especially in particle physics, where entangled $K^0\bar{K}^0$ and $B^0\bar{B}^0$
pairs are considerable examples. If future investigations will confirm
the violation of Bell's inequalities,
it is clear that, under the philosophy of realism,
the locality assumption would be incompatible with experimental evidence. 
%Then, if this were the case, maintaining realism one should
%consider as a real fact of Nature a non--local behaviour of quantum phenomena.
This fact is not in conflict with the theory of relativity: actually, there is no
way to use quantum non--locality for faster--than--light communication.

In this talk we discuss the predictions of
local realistic schemes for a pair of correlated neutral kaons created in the decay 
of the $\phi$--meson. 
Unlike photons, kaons are detectable with high efficiency.
Moreover, for $K^0\bar{K}^0$ pairs, which can be copiously produced at a high
luminosity $\Phi$-factory, additional assumptions regarding detection
not implicit in local realism
(always implemented in the interpretation of experiments with
photon pairs \cite{Cl78}) are not necessary to derive Bell's inequalities \cite{Se90}.
A correlation experiment discriminating between local realism 
and quantum mechanics could be performed at the Frascati $\Phi$-factory \cite{Frasc}
in the future. 
%Indeed, being designed to measure direct $CP$ violation
%in the $K^0$--$\bar{K}^0$ system, such a factory employs high precision
%detectors. 

%*****************************************************************   
% EPR's argument and Local Realism
%*****************************************************************   
\section{EPR's argument and Local Realism for {\bf $\phi \to K^0\bar{K}^0$}}
\label{localrealism}
The starting point of EPR's argumentation was the following
condition for a \underline{complete theory}:
{\it every element of physical reality must have a counterpart in the physical theory}.
They defined the \underline{physical reality} by means of the following sufficient
criterion: {\it if, without in any way disturbing a system, we can predict with certainty
(i.e., with probability equal to unity) the value of a physical quantity, then there exists
an element of physical reality corresponding to this quantity}. 
In addition, for a system made of two correlated, spatially separated 
and non--interacting entities,
EPR introduced the following \underline{locality} assumption:
{\it since at the time of measurement the two systems no longer interact, no real
change can take place in the second system in consequence of anything that may
be done to the first system}. 
The previous criterion of reality supports the anthropocentric
point of view nowadays called {\it realism}: it asserts that quantum systems have
intrinsic and well defined properties even when they are not subject to
measurements. 

To exemplify EPR's argumentation, consider the case 
(EPR--Bohm's gedanken experiment \cite{Bm51}) of a particle with
total angular momentum zero which decays, at rest, into two spin $1/2$ particles,
1 and 2, which fly apart with opposite momenta. After a certain time,
when they do not interact any more,
the normalized spin wave function of the global system, 
which does not depend on the quantization direction of the spin, is:
\beq
\label{spin0}
|S=0,S_z=0\rangle=\frac{1}{\sqrt{2}}\left[|+\rangle_1|-\rangle_2
-|-\rangle_1|+\rangle_2\right] .
\eeq
For particles 1 and 2, $|+\rangle$ and $|-\rangle$ represent spin--up and spin--down states,
respectively, along a direction chosen as $z$--axis. 
A measurement of the spin component of particle 1 along $z$ produces a given outcome 
[which is not predetermined by the quantum state (\ref{spin0})] and
forces, {\it immediately}, the spin of particle 2 along the opposite direction.
Following EPR, the spin component
of particle 2 is an element of physical reality, since it can be predicted 
{\it with certainty} and {\it without in any way disturbing} particle 2. Moreover, 
in order to fulfil the locality assumption (no action--at--a--distance),
EPR assumed that such an element of reality existed independently of any measurement
performed on particle 1. 

Following EPR's argumentation, 
the interpretation of the above experiment by means of quantum mechanics
leads to a difficulty. In fact, if we had performed a measurement of the spin component of 
particle 1 along another direction, say along the $x$-axis, this would have defined 
the $x$ component of the spin of particle 2 as another element of reality,
again independent of measurement. Obviously, this is also
valid for any spin component; then it should be possible, in the supposed complete
theory, to assign different spin wave functions
to the same physical reality. Therefore, one arrives 
at the conclusion that two or more physical
quantities which correspond to non--commuting quantum 
operators, can have simultaneous reality.
However, this is not possible in quantum mechanics.
Therefore, there exist elements of physical reality for which quantum 
mechanics has no counterpart, and, according to EPR's completeness definition, 
quantum theory cannot give a complete description of reality.

Actually, one could object, with Bohr \cite{Bo35}, that
in connection with a correlated system of non--interacting
subsystem, EPR's reality criterion
reveals the following weak point: it is not correct to assert that the measurement on
subsystem 1 does not disturb system 2; in fact, in quantum mechanics
the measurement do separate systems
1 and 2, which are not separated entities before the reduction of the wave
packet. 
%Moreover, in quantum mechanics
%two or more physical quantities can be considered as simultaneous elements
%of reality only when they can be simultaneously measured. 
Then, from the point of view
of orthodox quantum mechanics, EPR's argumentation ceases to be a paradox:
EPR's proof of incompleteness is mathematically correct 
but is founded on premises which are
inapplicable to microphenomena. 

%*****************************************************************
% Local Realism for the two--neutral--kaon system
%*****************************************************************
%\subsection{Local Realism for the two--neutral--kaon system}
%\label{localrealismkappa}
Now we come to the neutral--kaon system. In the following discussion
we shall neglect the (small) effects
of $CP$ violation. Then, the $CP$ eigenstates are identified with the short and
long living kaons (mass eigenstates): $|K_+ \rangle \equiv |K_S \rangle$ ($CP=+1$),
$|K_- \rangle \equiv |K_L \rangle$ ($CP=-1$). In this approximation the strong
interaction eigenstates $|K^0 \rangle$ and $|\bar{K}^0 \rangle$ are given by:
\beq
\label{s-eig}
|K^0 \rangle =  \frac{1}{\sqrt 2}\left[|K_S\rangle + |K_L\rangle\right] ,
\hspace{0.7cm}   
|\bar{K}^0 \rangle =  \frac{1}{\sqrt 2}\left[|K_S\rangle - |K_L\rangle\right] .
\eeq
The time evolution of the mass eigenstates is:
\beq
\label{time}
|K_{S,L}(\tau) \rangle=e^{-i\lambda_{S,L}\tau}|K_{S,L}\rangle ,
\eeq
where $|K_{S,L}\rangle\equiv |K_{S,L}(0)\rangle$,
$\tau= t \sqrt{1-v^2}$ is the kaon proper time 
[$t$ ($v$) being the time (kaon velocity) measured in the laboratory frame] and:
\beq
\lambda_{S,L}=m_{S,L}-\frac{i}{2}\Gamma_{S,L} ,
\eeq
$m_{S,L}$ denoting the $K_S$ and $K_L$ masses and $\Gamma_{S,L}$ the
corresponding decay widths: $\Gamma_{S,L}\equiv 1/{\tau_{S,L}}$
(we use natural units: $\hbar =c=1$).

Consider now the strong decay of the $J^{PC}=1^{--}$ $
\phi(1020)$--meson into $K^0\bar{K}^0$.
Just after the decay, at proper time $\tau=0$,
the quantum--mechanical state is given by the following superpositions:
\beq
\label{kkk}
|\phi(0)\rangle  =  \frac{1}{\sqrt 2}\left[  
|K^0\rangle_1 |\bar{K}^0\rangle_2 - |\bar{K}^0\rangle_1 |K^0\rangle_2\right]
=  \frac{1}{\sqrt 2}\left[
|K_L\rangle_1 |K_S\rangle_2 - |K_S\rangle_1 |K_L\rangle_2\right] .
\eeq
From eqs.~(\ref{s-eig}) and
(\ref{time}) the time evolution of state (\ref{kkk}) is obtained in the
following form:
\begin{eqnarray}
\label{qm2k}
|\phi(\tau_1,\tau_2)\rangle & = & \frac{1}{\sqrt 2}\left\{
e^{-i(\lambda_L\tau_1+\lambda_S\tau_2)}|K_L\rangle_1|K_S\rangle_2
-e^{-i(\lambda_S\tau_1+\lambda_L\tau_2)}|K_S\rangle_1|K_L\rangle_2\right\} \\ \nonumber
& = & \frac{1}{2\sqrt 2}\left\{
\left[e^{-i(\lambda_L\tau_1+\lambda_S\tau_2)}+
e^{-i(\lambda_S\tau_1+\lambda_L\tau_2)}\right]
\left[|K^0\rangle_1 |\bar{K}^0\rangle_2 - |\bar{K}^0\rangle_1 |K^0\rangle_2\right]\right. \\
\nonumber
& & \left. +\left[e^{-i(\lambda_L\tau_1+\lambda_S\tau_2)}-
e^{-i(\lambda_S\tau_1+\lambda_L\tau_2)}\right]
\left[|K^0\rangle_1 |K^0\rangle_2 - |\bar{K}^0\rangle_1 |\bar{K}^0\rangle_2\right]
\right\} .
\end{eqnarray}

Therefore, quantum mechanics predicts (and we know it is a well tested property)
a perfect anti--correlation in strangeness and $CP$ values
when both kaons are considered at the same time.
In the case in which both kaons are undecayed,
if an experimenter observes, say along direction 1, a $K^0$ ($K_L$), at
the same time
$\tau_1$, along direction 2, because of the instantaneous collapse of the two--kaon wave function, 
one can predict the presence of a $\bar{K}^0$ ($K_S$). Thus, at time $\tau_1$ to the kaon 
moving along direction 2 we assign an element of reality
(since, following EPR's reality criterion, the value of the corresponding physical quantity
is predicted {\it with certainty} and {\it without in any way disturbing}
the system), the value $-1$ ($+1$) of strangeness ($CP$). 
The same discussion is valid when the state observed
along direction 1 is $\bar{K}^0$ (or $K_S$) as well as when one exchanges the kaon
directions: $1\leftrightarrow2$. For times $\tau_2$ following the observation at time
$\tau_1$ along direction 1 of a $K_L$ ($K_S$), a $CP$
measurement on the other kaon will give with certainty the same result
$CP=+1$ ($CP=-1$) one expects at time $\tau_1$. This expresses $CP$ conservation.
In the case in which neither kaon has decayed, when the kaon detected at time $\tau_1$
is $K^0$ ($\bar{K}^0$), at times $\tau_2>\tau_1$ along direction 2 quantum
mechanics predicts the possibility to observe a 
$\bar{K}^0$ ($K^0$) as well as a $K^0$ ($\bar{K}^0$):
since strangeness is not conserved during the evolution
of the system, perfect anti--correlation on strangeness only exists
when both particles are considered at the same time. 

Following EPR's argument, 
in a local realistic approach one then associates to both kaons of the pair,
at any time, two elements of reality, which are not created by
measurements eventually performed on the partner when the
particles are space--like separated (locality): one determines the 
kaon $CP$ value, the other one supplies the kaon strangeness $S$.
They are both well defined also when the meson is not observed (realism)
and can take two values, $\pm 1$, which appear at random with the same 
frequency in a statistical ensemble of kaons. 
Therefore, four different single kaon states can
appear just after the $\phi$ decay, with the same
frequency (25\%). They are quoted in table \ref{loc-rea}. 
\begin{table}[t]
\begin{center}
\caption{Kaon realistic states.}
\label{loc-rea}
\begin{tabular}{c|c c}
\mc {1}{c|}{State} &
\mc {1}{c}{Strangeness} &
\mc {1}{c}{CP} \\ \hline
$K_1\equiv K^0_{S}$      & $+1$& $+1$  \\
$K_2\equiv \bar{K}^0_S$  & $-1$& $+1$  \\
$K_3\equiv K^0_L$        & $+1$& $-1$  \\
$K_4\equiv \bar{K}^0_L $ & $-1$& $-1$  \\
\end{tabular}
\end{center}
\end{table}
It is clear that this classification is incompatible with
quantum mechanics, where strangeness and $CP$
cannot be measured simultaneously.

%********************************************************************
\section{Quantum--mechanical evolution}
%********************************************************************
\label{qmp}

By introducing the shorthand notation:
\beq
E_{S,L}(\tau)=e^{-\Gamma_{S,L}\tau} , \hspace{0.7cm}
\Delta m=m_L-m_S ,  
\eeq
from eq.~(\ref{qm2k}) the quantum--mechanical (QM) probability 
$P_{QM}[K^0(\tau_1),\bar{K}^0(\tau_2)]\equiv |_1\langle K^0|_2\langle \bar{K}^0|
\phi(\tau_1,\tau_2)\rangle|^2$ that a measurement detects a $K^0$ at time 
$\tau_1$ along direction $1$ and a $\bar{K}^0$ at time
$\tau_2$ along direction $2$ is:
\begin{eqnarray}
\label{kkb}
P_{QM}[K^0(\tau_1),\bar{K}^0(\tau_2)]& =& P_{QM}[\bar{K}^0(\tau_1),K^0(\tau_2)] \\
& =&  \frac{1}{8}\left[E_L(\tau_1)E_S(\tau_2)+E_S(\tau_1)E_L(\tau_2)\right]
\left[1+A_{QM}(\tau_1,\tau_2)\right] .
\nonumber
\end{eqnarray}
The other probabilities relevant for our discussion are the following ones:
\begin{eqnarray}
\label{kk}
P_{QM}[K^0(\tau_1),K^0(\tau_2)]& =& P_{QM}[\bar{K}^0(\tau_1),\bar{K}^0(\tau_2)] \nonumber \\
& = & \frac{1}{8}\left[E_L(\tau_1)E_S(\tau_2)+E_S(\tau_1)E_L(\tau_2)\right]
\left[1-A_{QM}(\tau_1,\tau_2)\right] , \\
\label{ls}
P_{QM}[K_L(\tau_1),K_S(\tau_2)]&=&\frac{1}{2}E_L(\tau_1)E_S(\tau_2) , \\
\label{sl}
P_{QM}[K_S(\tau_1),K_L(\tau_2)]&=&\frac{1}{2}E_S(\tau_1)E_L(\tau_2) .
\end{eqnarray}
In eqs.~(\ref{kkb}) and (\ref{kk}):
\begin{eqnarray}
\label{qmasymm}
A_{QM}(\tau_1,\tau_2)&\equiv &\frac{P_{QM}[K^0(\tau_1),\bar{K}^0(\tau_2)]- 
P_{QM}[K^0(\tau_1),K^0(\tau_2)]} {P_{QM}[K^0(\tau_1),\bar{K}^0(\tau_2)]+ 
P_{QM}[K^0(\tau_1),K^0(\tau_2)]} \\
&=&2\frac{\sqrt{E_L(\tau_2-\tau_1)E_S(\tau_2-\tau_1)}}
{E_L(\tau_2-\tau_1)+E_S(\tau_2-\tau_1)}{\rm cos}\, \Delta m (\tau_2-\tau_1) , \nonumber
\end{eqnarray}
is the quantum--mechanical asymmetry parameter.

%%%%%%%%%%%%%%%%%%%%%%%%%%%%%%%%%%%%%%%%%%%%%%%%%%%%%%%%%%%%%%%%%%%%%%%%$$$$$$$$$$$$$$
\section{Local realistic evolution}
%%%%%%%%%%%%%%%%%%%%%%%%%%%%%%%%%%%%%%%%%%%%%%%%%%%%%%%%%%%%%%%%%%%%%%%%$$$$$$$$$$$$$$
\label{lrp}

In this section we discuss the predictions of 
local hidden--variable models for the kaon--pair observables.
More details can be found in ref.~\cite{DG00}. 

The quantum--mechanical expectation values for the evolution of a single kaon
can be reproduced by a realistic approach \cite{Se97}. 
Consider now the time evolution of a kaon pair in $\phi \to K^0\bar{K}^0$. 
At time $\tau =0$, immediately after the $\phi$ decay, in the realistic
picture there are four possible states for the pair, each appearing 
with a probability equal to $1/4$: they are listed in table~\ref{corr0}.
\begin{table}[t]
\begin{center}
\caption{Realistic states for the kaon pair at initial time $\tau =0$.}
\label{corr0}   
\begin{tabular}{c c}
\mc {1}{c}{Direction 1} &
\mc {1}{c}{Direction 2} \\ \hline
$K_1\equiv K^0_{S}$ \hspace{0.4cm}       ($S=+1$, $CP=+1$)  &   
$K_4\equiv \bar{K}^0_{L}$ \hspace{0.4cm} ($S=-1$, $CP=-1$) \\
$K_2\equiv \bar{K}^0_S$ \hspace{0.4cm}   ($S=-1$, $CP=+1$)  & 
$K_3\equiv K^0_L$ \hspace{0.4cm}         ($S=+1$, $CP=-1$) \\
$K_3\equiv K^0_L$ \hspace{0.4cm}         ($S=+1$, $CP=-1$)  & 
$K_2\equiv \bar{K}^0_S$ \hspace{0.4cm}   ($S=-1$, $CP=+1$)  \\
$K_4\equiv \bar{K}^0_L$ \hspace{0.4cm}   ($S=-1$, $CP=-1$)  &
$K_1\equiv K^0_S$ \hspace{0.4cm}         ($S=+1$, $CP=+1$)   \\
\end{tabular}
\end{center}
\end{table}
We assume, as in quantum mechanics, 
a perfect anti--correlation in strangeness and $CP$
when both kaons are considered at equal times. 

When the system evolves, the kaons fly apart from each other,
and at two generic times $\tau_1$ and $\tau_2$
(corresponding to opposite directions of propagation labeled 1 and 2, respectively)
the kaon pair is in one of the states reported in table~\ref{corrt}.
\begin{table}[t]
\begin{center}
\caption{Local realistic states for the kaon pair at times $\tau_2\geq \tau_1$.}
\label{corrt}
\begin{tabular}{l | c c}
\mc {1}{c |}{Probabilities}&
\mc {1}{c}{Direction 1 (Left) \hspace{0.2cm}Time $\tau_1$} &
\mc {1}{c}{Direction 2 (Right)\hspace{0.2cm}Time $\tau_2$} \\ \hline
$P_1(\tau_1,\tau_2;\lambda)$ & $K_1\equiv K^0_{S}$      &  $K_4\equiv \bar{K}^0_L$  \\
$P_2(\tau_1,\tau_2;\lambda)$ & $K_1\equiv K^0_{S}$      &  $CP=-1$ DP      \\  
$P_3(\tau_1,\tau_2;\lambda)$ & $CP=+1$ DP     &   $K_4\equiv \bar{K}^0_L$  \\
$P_4(\tau_1,\tau_2;\lambda)$ & $K_1\equiv K^0_{S}$      &  $K_3\equiv K^0_L$  \\
. . . . . . .& . . . . & . . . . \\
\end{tabular}
\end{center}
\end{table}
The first row refers to the state with a $K_1$ at time $\tau_1$ along direction
$1$ (left) and a $K_4$ at time $\tau_2$ along direction $2$ (right).
Given the classification of the table, 
in our discussion we consider $\tau_2\geq\tau_1$: the isotropy of space
guarantees the invariance of the two--kaon states by exchanging the directions 1 and 2.
In the second row the state corresponds to a left going $K_1$ at time $\tau_1$ 
and $CP=-1$ decay products (DP) at time $\tau_2$ on the right. 
These decay products originate from the instability
of the $K_3$ and $K_4$ pure states, which are both long living kaons, namely
$CP=-1$ states.
%(the corresponding physical processes are: 
%$K_L\to 3\pi, \pi \mu \nu_{\mu}, \pi e \nu_e$). 
At time $\tau_1$ the state correlated
with a left going $K_1$ is necessarily either a $K_4$ or a state containing 
$CP=-1$ decay products, $K^{DP}_3$ or $K^{DP}_4$. 
Then, at time $\tau_2$ ($> \tau_1$) on the right we can have:
i) a $K_4$ (state in the first row), ii) $CP=-1$ decay products (state in
the second row) or iii) a $K_3$ (state in the fourth row). The former case refers to the
transition $K_4(\tau_1)\to K_4(\tau_2)$, the latter to
$K_4(\tau_1)\to K_3(\tau_2)$, both along direction 2. 
Occurrence ii) takes contributions
from the following transitions: $K^{DP}_3(\tau_1)\to K^{DP}_3(\tau_2)$,
$K^{DP}_4(\tau_1)\to K^{DP}_4(\tau_2)$,
$K_4(\tau_1)\to K^{DP}_4(\tau_2)$ and $K_4(\tau_1)\to K_3(\tau_1<\tau<\tau_2)
\to K^{DP}_3(\tau_2)$. The other 14 local realistic states not quoted
in table~\ref{corrt} can be obtained in the same way \cite{DG00}.

It is important to stress that the states listed in 
table~\ref{corrt} are assumed to be well defined
for all times $\tau_1$ and $\tau_2$ with $\tau_1\leq \tau_2$: 
this is the main requirement of the 
realistic approach. For a given kaon pair, in a deterministic theory
only one of the possibilities of table~\ref{corrt} really occurs
for fixed $\tau_1$ and $\tau_2$. This means that we are making the
hypothesis (realism) that there exist additional variables $\lambda$, called
{\it hidden--variables}, that provide
a complete description of the pair, which is viewed as really existing
and with well defined properties independently of any observation. 
The state representing the meson pair for given times $(\tau_1,\tau_2)$
is completely defined by these hidden--variables:
they are supposed to determine in advance (say when
the two kaons are created) the future behaviour of the pair. Thus,
the times in correspondence of which the instantaneous $|\Delta S|=2$ jumps and the decay
occur for a given kaon are predetermined by its hidden--variables.
Under this hypotheses there is no problem concerning a possible causal influence
acting among the different entities of entangled systems when a measurement
takes place on one subsystem. However, the new variables are 
unobservable since they are averaged out in the measuring processes, and  
unobservable are the states of table~\ref{corrt}. Besides, 
1) also the measuring apparata could be
described by means of hidden--variables, which influence the results of measurement,
and 2) hidden--variables associated to the kaon pair could show a non--deterministic
behaviour. For further details concerning the hidden--variable
interpretation of the states in table \ref{corrt} see ref.~\cite{DG00}.

In ref.~\cite{DG00} we have studied the range of variability of the meson
pair observables compatible with the most general local realistic
model, obtaining the following inequality on the asymmetry parameter:
\beq
\label{interval}  
2|Q_+(\tau_2)-Q_-(\tau_1)|-1\leq A_{LR}(\tau_1,\tau_2)\leq 
1-2|Q_+(\tau_2)-Q_+(\tau_1)| ,
\eeq
where:
\beq
Q_{\pm}(\tau)=\frac{1}{2}\left[1\pm 2\frac{\sqrt{E_L(\tau)E_S(\tau)}} 
{E_L(\tau)+E_S(\tau)}{\rm cos}\,\Delta m \tau \right]  .
\eeq

%*****************************************************************
% Compatibility between local realism and quantum mechanics
%*****************************************************************
\section{Compatibility between local realism and quantum mechanics}
\label{comp}

Local realism reproduces the quantum--mechanical predictions
for the single kaon observables and the joint probabilities 
(\ref{ls}), (\ref{sl}). 
The same conclusion would be true for the
observables (\ref{kkb}) and (\ref{kk})
involving $K_S$--$K_L$ mixing if the time--dependent local realistic 
asymmetry parameter had the same expression it has in quantum mechanics:
\beq
\label{lr-eq-qm}
{\rm Local\; Realism\; equivalent\; to\; Quantum\; Mechanics}\hspace{0.2cm}
\Longleftrightarrow
\hspace{0.2cm}A_{LR}(\tau_1,\tau_2)\equiv A_{QM}(\tau_1,\tau_2) .
\eeq

\begin{figure}[thb]
\begin{center}
\input{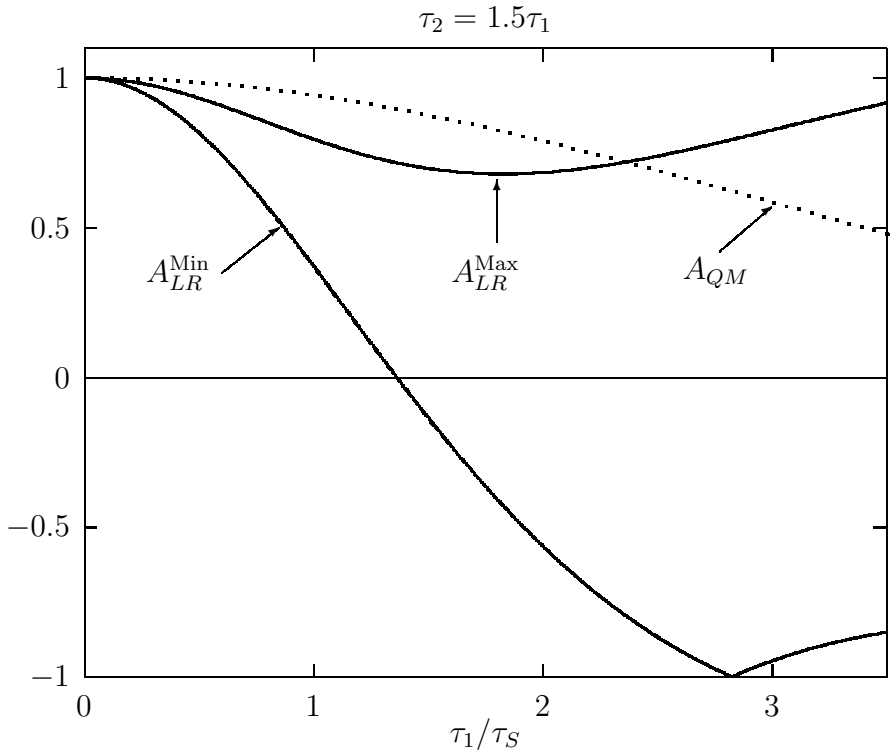}
\end{center}
%\mbox{\epsfig{file=asimmetria1_5.eps,width=.8\textwidth}}
\vskip 1.5mm
\caption{Local realistic and quantum--mechanical asymmetry parameters for $\tau_2=1.5\tau_1$
plotted vs $\tau_1/\tau_S$.}
\label{inc1}
\end{figure}

From eq.~(\ref{interval}) it follows that for
the special case of $\tau_2=\tau_1\equiv \tau$, local realism
is compatible with quantum mechanics. This also occurs when 
$\tau_1\equiv 0$.
Another special case is when, for instance, $\tau_2=1.5\tau_1$: in this situation, 
the local realistic asymmetry does not satisfy
the compatibility requirement (\ref{lr-eq-qm}). This is depicted in figure~\ref{inc1}. 
There is an evident discrepance when $0< \tau_1\lsim 2.3\tau_S$,
the largest incompatibility corresponding to $\tau_1\simeq 1.5\tau_S$,
where $[A_{QM}-A_{LR}^{\rm Max}]/A_{QM}\simeq 20$\%. 
In general, local realism and quantum mechanics are incompatible 
when $\tau_2=\alpha\tau_1$ with $\alpha> 1$. The degree of incompatibility
increases for increasing $\alpha$. For instance, when $\tau_2=2\tau_1\simeq 2.4\tau_S$, 
$A_{QM}$ is 27~\% larger than $A_{LR}^{\rm Max}$.

However, it is important to stress the following restriction concerning the 
choice of the detection times $\tau_1$ and $\tau_2$.
In order to satisfy the locality condition, namely to make sure that the measurement
on the right is causally disconnected from that on the left,
these events must be space--like separated. In the center of mass of 
the process $\phi \to K^0\bar{K}^0$, this requirement corresponds to choose 
detection times in the interval: $1\leq \tau_2/ \tau_1< 1.55$.

An experiment that measured the asymmetry parameter
was performed by the CPLEAR collaboration at CERN \cite{Ap98}.
The $K^0\bar{K}^0$ pairs were produced by proton--antiproton
annihilation at rest. Unfortunately, because of their large error bars,
the CPLEAR data are in agreement, within one
standard deviation, with both quantum mechanics and local realism.

%***************************************************************** 
% BELL´s INEQS
%*****************************************************************
\section{Bell's inequalities} 
\label{imposs} 

Because of the particular values of the kaon lifetimes ($\Gamma_S$ and $\Gamma_L$) 
and of the quantity $\Delta m\equiv m_L-m_S$, it is impossible
to show a violation, by quantum mechanics,
of Bell's inequalities exploiting strangeness measurements
at different times.
This was the conclusion of ref.~\cite{Gh91}. 
In this section we consider again this question in order 
to show how a Bell--type test is actually feasible.
The reason of the difficulty in designing a Bell test with kaons 
lies in the very short $K_S$ lifetime ($\tau_S$) compared with the typical time 
($2\pi/\Delta m\simeq 13 \tau_S$) of the strangeness oscillations. 

Consider joint probabilities normalized to undecayed kaon pairs:
\beq
\label{replace-prob}
P[\bar{K}^0(\tau),\bar{K}^0(\tau')]\to 
P^{\rm ren}[\bar{K}^0(\tau),\bar{K}^0(\tau')]\equiv
\frac{P[\bar{K}^0(\tau),\bar{K}^0(\tau')]}{P[-(\tau),-(\tau')]} \\
=\frac{1}{4}[1-A(\tau,\tau')] , 
\eeq
where the probability that at times $\tau$ (on the left)
and $\tau'$ (on the right) the kaons are undecayed is: 
\beq
P[-(\tau),-(\tau')]=
\frac{1}{2}[E_S(\tau)E_L(\tau')+E_L(\tau)E_S(\tau')] ,
\eeq
both in the local realistic description and in quantum mechanics.
The renormalized observables are less damped than the original ones, and,
as a consequence, a Bell--type test can be performed.

The same derivation that supplies Clauser--Horne--Shimony--Holt's
(CHSH's) inequality \cite{CHSH69,CH74} in the standard (namely
unrenormalized) case can be applied to the renormalized
observables of eq.~(\ref{replace-prob}). By introducing four detection times
($\tau_1$ and $\tau_2$ for the left going meson, 
$\tau_3$ and $\tau_4$ for the right going meson),
CHSH's inequality for strangeness $-1$ detection is then:
\beq
\label{chsh}
-1\leq S_{LR}(\tau_1,\tau_2,\tau_3,\tau_4)\leq 0 ,
\eeq
with:
\begin{eqnarray}
\label{chsh1}
S_{LR}(\tau_1,\tau_2,\tau_3,\tau_4)&\equiv&
P_{LR}^{\rm ren}[\bar{K}^0(\tau_1),\bar{K}^0(\tau_3)]-
P_{LR}^{\rm ren}[\bar{K}^0(\tau_1),\bar{K}^0(\tau_4)] 
+P_{LR}^{\rm ren}[\bar{K}^0(\tau_2),\bar{K}^0(\tau_3)] \\
&&+P_{LR}^{\rm ren}[\bar{K}^0(\tau_2),\bar{K}^0(\tau_4)]-
P_{LR}^{\rm ren}[\bar{K}^0(\tau_2)]-P_{LR}^{\rm ren}[\bar{K}^0(\tau_3)] , \nonumber
\end{eqnarray}
where $P_{LR}^{\rm ren}[\bar{K}^0(\tau)]\equiv P_{LR}[\bar{K}^0(\tau)]/
P_{LR}[-(\tau)]=1/2$.
Consider the special case in which the four times are related by:
\beq
\label{timerel}
\tau_3-\tau_1=\tau_2-\tau_3=\tau_4-\tau_2=\frac{1}{3}(\tau_4-\tau_1)\equiv \tau .
\eeq
Thus, in quantum mechanics eq.~(\ref{chsh1}) reduces to:
\beq
\label{chsh2}
S_{QM}(\tau)=\frac{1}{4}\left[2-3A_{QM}(\tau)+A_{QM}(3\tau)\right]-1 . 
\eeq

\begin{figure}[t]
\begin{center}
\input{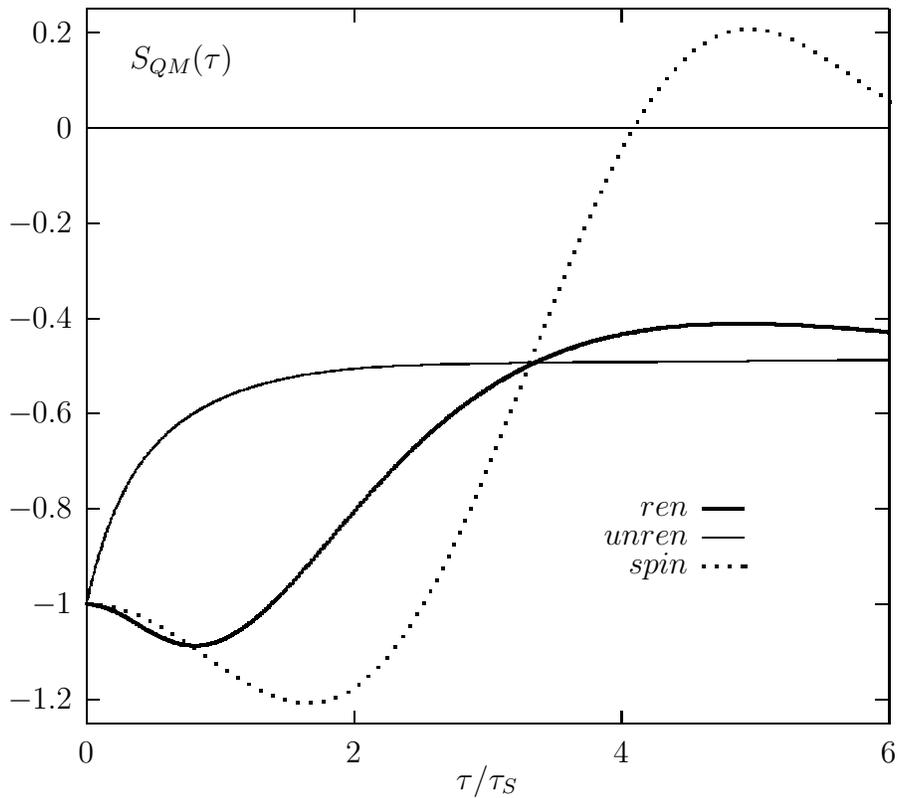}
\end{center}
%\mbox{\epsfig{file=chsh.eps,width=.8\textwidth}}
\vskip 1.5mm
\caption{Violation of CHSH's inequality (\ref{chsh})
for $\tau_1/p=\tau_2/(p+2)=\tau_3/(p+1)=\tau_4/(p+3)\equiv \tau$.
The function $S_{QM}$ of eq.~(\ref{chsh2}) is plotted versus $\tau$.
See text for further details.}
\label{chshfig}
%\end{center}
\end{figure}

If we choose $\tau_1\equiv \tau$, the other times become:
$\tau_2=3\tau$, $\tau_3=2\tau$ and $\tau_4=4\tau$, and,
in the limit of stable kaons ($\Gamma_S=\Gamma_L=0$), both side of inequality
(\ref{chsh}) are violated by quantum mechanics in periodical intervals of $\tau$
(see curve marked {\it spin} in figure \ref{chshfig}): this situation
correspond to the case of the spin--singlet system system (\ref{spin0}).

As far as the real case for kaons is considered, quantum mechanics does not
violate inequality (\ref{chsh}) when unrenormalized expectation values
are used (see curve {\it unren} in figure \ref{chshfig}). The conclusion is
different once one employs probabilities normalized 
to undecayed kaon pairs: as it is shown
in figure \ref{chshfig} (curve {\it ren}), for $0< \tau\lsim 1.4\tau_S$
quantum--mechanical expectation values are incompatible with the left hand side of 
inequality (\ref{chsh}). The largest violation of the inequality 
($-1.087<-1$) corresponds to $\tau\simeq 0.81\tau_S$ and 
$P_{QM}^{\rm ren}[\bar{K}^0(\tau_1),\bar{K}^0(\tau_3)]\simeq 0.036$,
$P_{QM}^{\rm ren}[\bar{K}^0(\tau_1),\bar{K}^0(\tau_4)]\simeq 0.195$.

With the previous choice of the four detection times the locality condition 
$1\leq \tau_2/\tau_1< 1.55$ 
is not satisfied, since: $\tau_4/\tau_1=4>$. In order to
fulfil this requirement when relation (\ref{timerel}) is used, one can introduce
times $\tau_1=p\tau$, $\tau_2=(p+2)\tau$, $\tau_3=(p+1)\tau$ and $\tau_4=(p+3)\tau$
($p\geq 0$) and require $\tau_4/\tau_1=(p+3)/p< 1.55$, thus $p> 5.45$. 
However, since the renormalized quantum--mechanical probabilities
only depend on the difference between the observation times
[see eqs.~(\ref{replace-prob}), (\ref{qmasymm})],
the result {\it ren} of figure \ref{chshfig}
is independent of $p$, and the locality condition is satisfied.
Thus, experimentally one could choose to use, for instance,
$p=6$, namely $\tau_1=6\tau$, $\tau_2=8\tau$, $\tau_1=7\tau$, 
$\tau_4=9\tau$, and the largest violation of the inequality would be again
for $\tau\simeq 0.81\tau_S$. However, as $p$ increases, even if the renormalized
probabilities are unchanged, the strangeness
detection becomes more and more difficult, because of the kaon decays, thus 
small $p$ are preferable.
Also the curve corresponding to the limit $\Gamma_S=\Gamma_L=0$ is the same for any $p$.
The curve corresponding to the inequality
that makes use of unrenormalized probabilities depends on $p$, but this case 
is not interesting since it cannot be used
for a discriminating test whatever the choice of $p$ is. 

%*****************************************************************
% Conclusions
%*****************************************************************
\section{Conclusions}
\label{concl}
In agreement with Bell's theorem, in this talk we have shown that
quantum mechanics for the two--neutral--kaon system cannot be completed by a
theory which is both local and realistic: the separability assumed 
in Bell's local realistic theories for the joint probabilities contradicts the
non--separability of quantum entangled states. 
Any local realistic approach is only able to reproduce the non--paradoxical
predictions of quantum mechanics like the
perfect anti--correlations in strangeness and $CP$ and the single 
particle observables.

The incompatibility proof among quantum mechanics and local realistic models
has been carried out by employing two different
approaches. We started discussing the variability of the $K^0\bar{K}^0$
expectation values deduced from the general premises
concerning locality and realism. The realistic states have been interpreted
within the widest class of hidden--variable models. 
%As far as the process $e^+e^-\to \phi \to K^0\bar{K}^0$ is considered,
Under particular conditions
for the experimental parameters (the detection times), the discrepancies 
among quantum mechanics and local realistic models for the 
time--dependent asymmetry parameter are not less than
20\%. The data collected by the CPLEAR collaboration for the asymmetry 
%(which used the reaction $p \bar{p}\to K^0\bar{K}^0$ ) 
do not allow for conclusive answers concerning a refutation of
local realism. 

The other approach that we followed makes use of Bell--like inequalities
involving $K_S$--$K_L$ mixing.
Contrary to what is generally believed in the literature, we have shown that a Bell--type
test is feasible at a $\Phi$--factory, using CHSH's inequalities. 

Concluding, by employing an experimental accuracy for joint kaon detection
considerably higher than that corresponding to the CPLEAR measurement,
a decisive test of local realism vs quantum mechanics
both with and without the use of Bell's inequalities will be feasible in the
future at the Frascati $\Phi$--factory.

%\acknowledgments
%We are grateful to Albert Bramon for many valuable discussions
%We also ackowledge the hospitality of Prof. D. Sherrington 
%of the Department of Theoretical Physics of Oxford University. 
%One of us (G.G.) acknowledge financial support by the EEC through 
%TMR Contract CEE--0169. 

\vfill\eject
\end{document}